\documentclass[reprint,
superscriptaddress,
amsmath,amssymb,
aps,dvipsnames,usenames]{revtex4-2}

\usepackage[utf8]{inputenc}
\usepackage[detect-all]{siunitx}
\usepackage{graphicx}

\begin{document}

 
  \title{Arbitrary polarization conversion for pure vortex generation with a single metasurface}
 \author{Marco Piccardo}
  \email{marco.piccardo@iit.it}
 \affiliation{Center for Nano Science and Technology,\\ Fondazione Istituto Italiano di Tecnologia, Milan 20133, Italy}

 \author{Antonio Ambrosio}
 \email{antonio.ambrosio@iit.it}
 \affiliation{Center for Nano Science and Technology,\\ Fondazione Istituto Italiano di Tecnologia, Milan 20133, Italy}
 

  
\begin{abstract}
      The purity of an optical vortex beam depends on the spread of its energy among different azimuthal and radial modes. The smaller is this spread, the higher is the vortex purity and the more efficient are its creation and detection. There are several methods to generate vortex beams with well-defined orbital angular momentum but only few exist allowing to select a pure radial mode. These typically consist of many optical elements with rather complex arrangements, including active cavity resonators. Here we show that it is possible to generate pure vortex beams using a single metasurface plate in combination with a polarizer. We generalize an existing theory of independent phase and amplitude control with birefringent nanopillars considering arbitrary input polarization states. The high purity, sizeable creation efficiency and impassable compactness make the presented approach a powerful complex amplitude modulation tool for pure vortex generation, even in the case of large topological charges.
\end{abstract}

\maketitle

\textit{Introduction.}---The characterizing feature of an optical vortex is a zero of intensity, which coincides with a phase singularity of the field. The phase circulates around this point of null intensity endowing the beam of light with orbital angular momentum \cite{Allen1992} (OAM). The OAM and the specific ring-shaped intensity distribution of these beams makes them attractive for a number of applications \cite{Franke-Arnold2020}, ranging from quantum information \cite{Mair2001,Nagali2009} to super-resolution microscopy \cite{Fuerhapter2005,Li2013}, and has motivated the development of several techniques of optical vortex generation \cite{Wang2018}. However, most of these methods, such as spiral phase plates \cite{Beijersbergen1994}, computer generated holograms \cite{Heckenberg1992,Ambrosio2012}, spatial light modulators \cite{Ngcobo2013}, $q$-plates \cite{Devlin2017a,Rubano2019} and $J$-plates \cite{Devlin2017,Huang2019}, rely on phase-only (PO) transformations. The azimuthal phase modulation imparted by these optical elements allows to create an optical vortex but the lack of amplitude modulation prevents the ouput beam from being a solution of the paraxial wave equation. The missing amplitude term is compensated by the spreading of the beam energy during propagation on high-order radial modes, thus leading to impure states consisting of a superposition of vortex modes \cite{Sephton:16}.

There exist a few methods that can provide both the required phase and amplitude (PA) modulation for pure vortex generation, such as mode conversion in active resonators \cite{Maguid2018,Uren2019,Sroor2020}, but these require either specific input beams or rather complex cavity set-ups. Here we introduce a method that allows to convert an arbitrary input beam into a pure vortex mode using a single metasurface plate, which is easy to implement in practical optics experiments and could be of use in any vortex application requiring that the beam power is contained within a specific radial mode.

\begin{figure*}[t]
    \centering
    \includegraphics[width=0.9\textwidth]{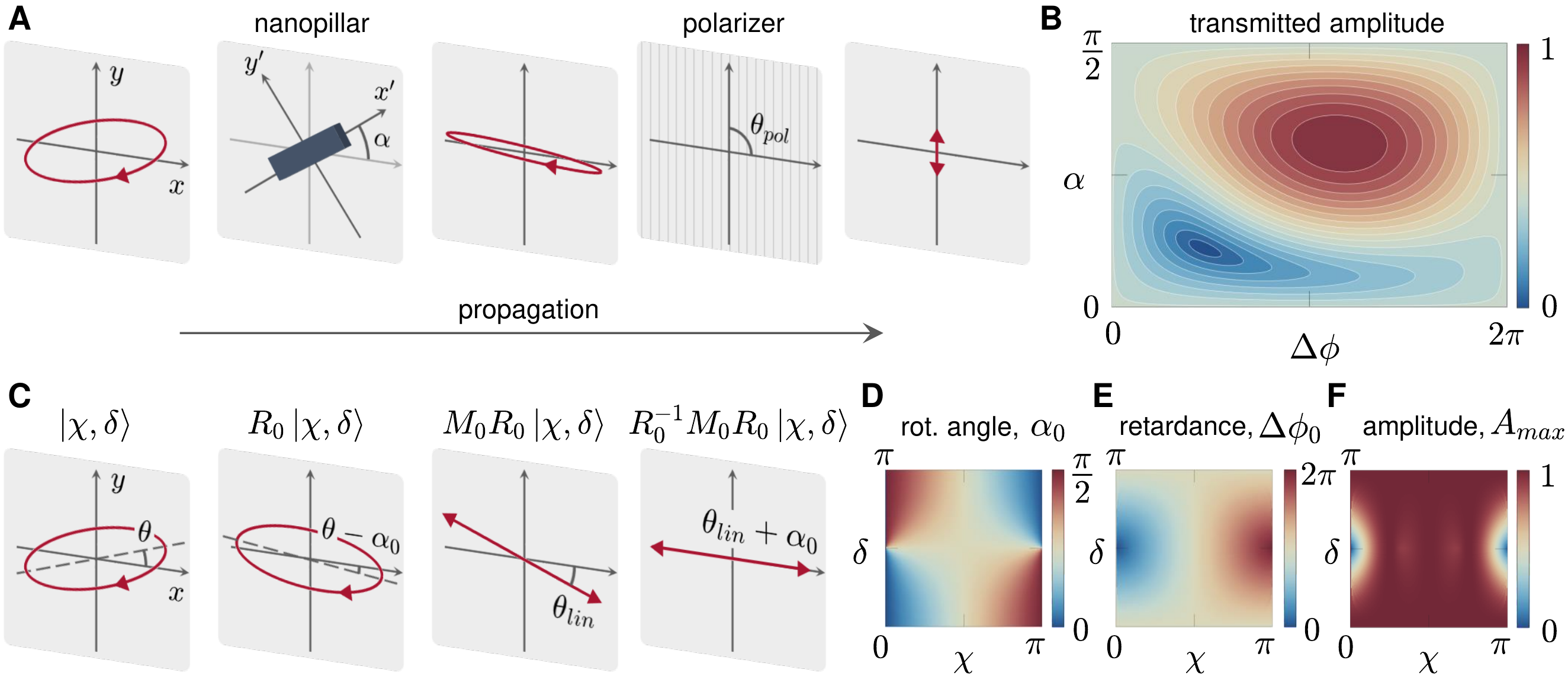}
    \caption{\textbf{Mapping polarization to amplitude.} \textbf{A}, An arbitrary input polarization state is converted by a birefringent nanopillar into a new polarization state that, after passing through a polarizer, is reduced in amplitude. This approach still allows to control independently the phase of the output state. The nanopillar orientation angle and polarizer axis are $\alpha$ and $\theta_{pol}$, respectively. \textbf{B}, Contour plot showing the transmitted amplitude of the elliptical state shown in \textbf{A} after propagation through the nanopillar--polarizer system as a function of the orientation angle $\alpha$ and phase retardance $\Delta\phi$ of the pillar. \textbf{C}, Conditions to achieve amplitude extinction through the nanopillar--polarizer system when the polarizer axis is aligned along the $y$-axis. The input is an arbitrary polarization state $\left| \chi, \; \delta \right>$ that is rotated by an angle $\alpha_0$ in the nanopillar frame of reference (operator $R_0$), linearized via birefringence (operator $M_0$), and finally brought back to the original frame of reference (operator $R_0^{-1}$) resulting orthogonal to the polarizer axis. \textbf{D,E}, Design parameter maps corresponding to the pillar angle and phase retardance needed to obtain extinction for any input polarization state, represented in the $(\chi,\delta)$ plane. \textbf{F}, Maximum amplitude transmission through the nanopillar--polarization system for any input state when the phase retardance is fixed to $\Delta\phi_0$ and only $\alpha$ is adjusted. If instead both parameters are allowed to vary a full modulation range from 0 to 1 can be achieved, as in \textbf{B}.}
    \label{fig_polarization_conversion}
\end{figure*}

\textit{Methods.}---After presenting our metasurface theory, we will prove the approach by applying it to the design of dielectric metasurfaces operating in the near-infrared. This spectral range is only chosen for the sake of illustration, without loss of generality. We consider amorphous silicon nanopillars with a rectangular cross section lying on a silica substrate. The pillars have a height of 600 nm and are arranged in an hexagonal close packed lattice (600 nm pillar--to--pillar separation). The wavelength of the source is 1064 nm. A library of the transmission coefficients and phases imparted by the pillars as a function of their size $L_{x,y}$ (in the 100--480 nm range) is constructed using the finite-difference time-domain (FDTD) module of Lumerical and selecting the elements with the best performance in terms of amplitude transmission and phase accuracy \cite{Arbabi2015}. The average transmittance of the pillars chosen from the library to design the metasurfaces considered in this work is typically around 95\%. The design code sets the sizes and orientation angles of the pillars according to our theory to produce a desired Laguerre-Gaussian (LG) beam. We validate the method by carrying out FDTD simulations of metasurfaces with 30 $\mu$m diameter, using plane-wave or Gaussian sources with linear or elliptical polarizations. The beam waist of the designed LG modes is set to $\omega_0 = 5\:\mu$m. The simulated near field is propagated to a hemispherical surface in the far field at 1 m from the metasurface and then projected to a plane for visual representation using direction cosines. Finally, the propagated field is decomposed on an LG basis to evaluate its modal purity. The basis is truncated at the 20$^{\mathrm{th}}$ radial mode. The complex beam parameter of the LG basis is obtained by an optimization algorithm that maximizes the overlap probability of the far field beam with its decomposition over the finite basis \cite{Vallone2017}.

\textit{Phase and amplitude control.}---The operation of a single birefringent nanopillar on an incident field can be represented in Jones calculus in the pillar frame of reference as
\begin{equation}
    M = \mathrm{e}^{i \psi} \left( \begin{matrix} 
                1 & 0 \\
                0  & \mathrm{e}^{i  \Delta\phi}
                \end{matrix} \right)
\label{eq_M}
\end{equation}
where the birefringence arises from the form factor of the pillar. We assume unity transmission, thus $M$ is unitary and the pillar cannot modulate the amplitude of the field. Meta-atom geometries capable of controlling directly both the phase and amplitude of the field have been demonstrated \cite{Wang2016,Jia2020} but suffer in general from fabrication complexities, as they rely on resonances, may be limited in efficiency, or may not allow full modulation ranges, i.e. from 0 to 1 in amplitude and from 0 to $2\pi$ in phase. However, a simple birefringent pillar represented by the operator $M$ allows to convert the polarization of the incident field thanks to the phase retardance $\Delta\phi$. Thus, by adding a linear polarizer after the metasurface one can in principle use the projection of the converted polarization state to modulate the transmitted amplitude of the field, as well as control its phase via the global $\mathrm{e}^{i \psi}$ factor. This simple but powerful scheme was demonstrated with metasurfaces designed for the specific cases of linearly-polarized \cite{Divitt2019} and circularly-polarized \cite{Overvig2019} light inputs. Here we generalize this approach to input states of arbitrary polarization.

The operation principle of the proposed scheme of PA control is shown in Fig. \ref{fig_polarization_conversion}A. Given an elliptically polarized input, the pillar dimensions and orientation angle $\alpha$ are designed so that the polarization-converted beam after passing through a polarizer has the desired phase and amplitude. The phase is determined by one of the two dimensions of the pillar, which sets $\psi$, while the amplitude can be controlled via a combination of $\Delta\phi$ and $\alpha$, as shown in Fig. \ref{fig_polarization_conversion}B.

\begin{figure*}[th!]
    \centering
    \includegraphics[width=0.9\textwidth]{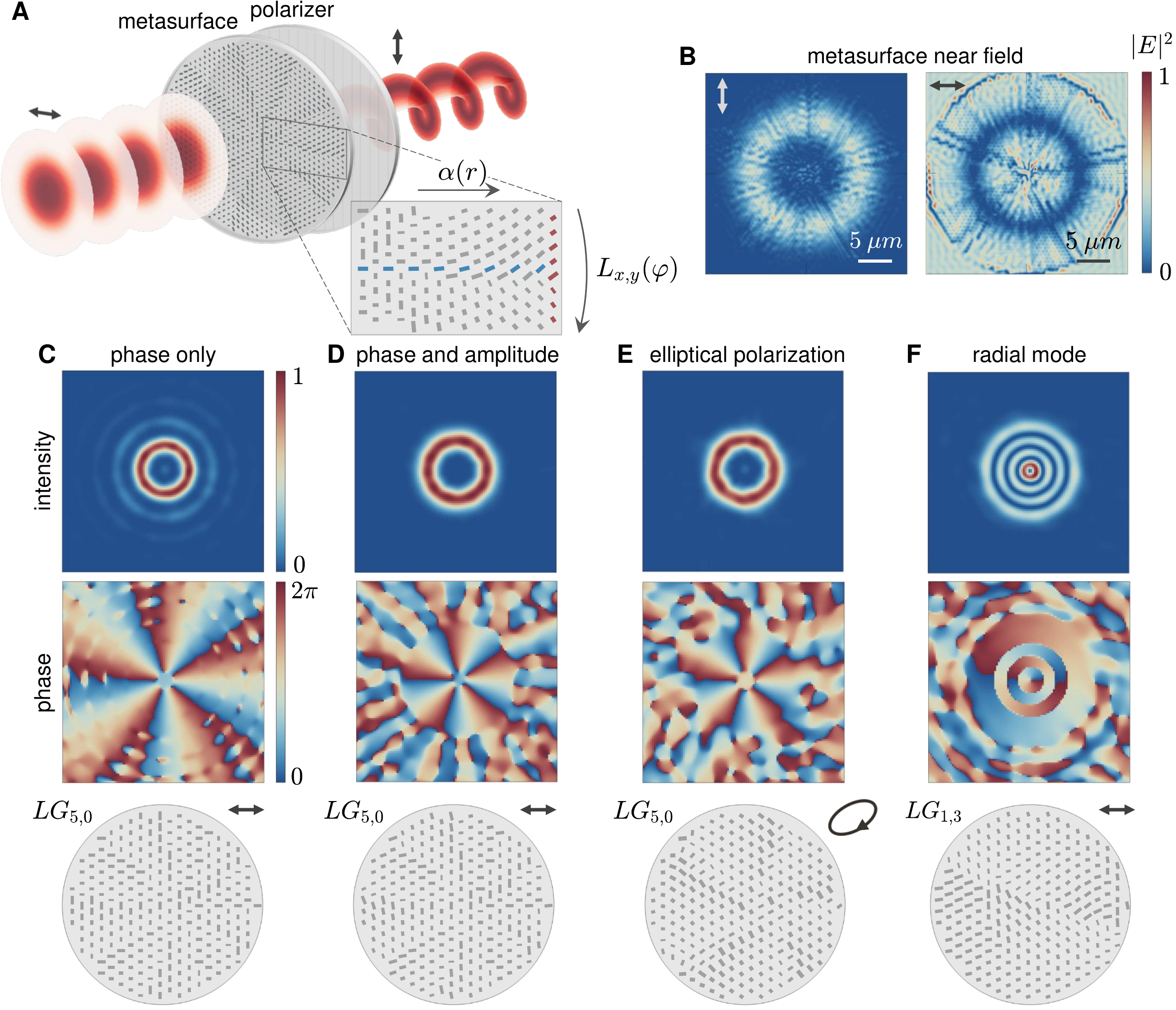}
    \caption{\textbf{Pure vortex generation from a single metasurface.} \textbf{A}, Schematic of the system implementation for pure vortex generation. An input wave of known amplitude, phase and polarization, here represented as a linearly-polarized Gaussian beam, propagates through the metasurface and polarizer and exits as a pure vortex beam. The angles of the nanopillars along the radial directions $\alpha(r)$ define the amplitude mask of the metasurface, while their rectangular dimensions set the azimuthal phase profile $L_{x,y}(\varphi)$ of the beam. \textbf{B}, Near-field intensity maps obtained by finite-difference time-domain (FDTD) simulations for a metasurface generating an $LG_{5,0}$ mode. Only the vertically-polarized component (left) is transmitted through the polarizer. \textbf{C--F}, Far field intensity and phase distributions obtained by FDTD simulations for different structures: \textbf{C}, a phase-only metasurface generating a superposition of radial modes; \textbf{D--F}, phase and amplitude metasurfaces generating pure vortex beams for different input and output states. In all cases the bottom row shows the central area (10 $\mu$m diameter) of the metasurface structures, the target Laguerre-Gaussian beam, and the input polarization state (either linear or elliptical).}
    \label{fig_pure_vortex_generation}
\end{figure*}

Now we concentrate on the specific problem of finding analytically the exact conditions to obtain the extinction. This is much more important than determining the conditions for unity transmission, as the extinction is critical to mask certain parts of the input beam. On the other hand the conditions for unity transmission only influence the efficiency of the device and will be discussed later. The problem is illustrated in Fig. \ref{fig_polarization_conversion}C. We consider an incident elliptical state represented by the Jones vector
\begin{equation}
    \left(
    \begin{matrix}
     \mathrm{cos}\chi \\ \mathrm{e}^{i\delta}\mathrm{sin}\chi 
    \end{matrix} \right)
    \equiv \left|\chi,\; \delta \right>
\end{equation}
and a linear polarizer with angle $\theta_{pol} = \pi/2$ with respect to the $x$-axis. Our goal is to find the pillar parameters that convert the input state into a linear state oriented perpendicularly to the polarizer axis. In the following the subscript ``0'' refers to parameters and operators corresponding to the extinction condition. First of all we use the rotation matrix $R_0$ to rotate the field on the pillar basis as $R_0 \left|\chi,\; \delta \right> \equiv \left|\chi',\; \delta' \right>$, which turns the ellipse angle from $\theta$ to $\theta-\alpha_0$. Then the new state can be linearized via the pillar birefringence operator as $M_0 \left|\chi',\; \delta' \right>$ with the condition $\Delta\phi_0 = -\delta'$. After linearization the orientation of the state is further rotated to an angle $\theta_{lin} = \chi'$. We impose that after returning back to the original reference frame using $R_0^{-1} M_0 R_0 \left|\chi,\; \delta \right>$ the output state lies parallel to the $x$-axis, which gives the condition $\theta_{lin} + \alpha_0 = 0$. By combining the previous conditions we obtain $\alpha_0 = -\chi'$. Thus all that remains is to express the rotated state $\left|\chi',\; \delta' \right>$ in terms of the input one $\left|\chi,\; \delta \right>$. This finally leads to the general extinction condition for the orientation angle
\begin{equation}
    \alpha_0 = \frac{1}{2}\;\mathrm{tan^{-1}}\left(\mathrm{tan}\:\chi \; \mathrm{sec}\:\delta  \right).
\end{equation}
valid for an arbitrary polarization state. A closed-form expression can also be deduced for the phase retardance $\Delta\phi_0$, though not as compact (Supplementary Material).

Fig. \ref{fig_polarization_conversion}D,E represent the solutions for the extinction condition for all the possible input polarization states shown on the $(\chi,\delta)$ plane. In the case of circularly-polarized light (i.e. $\left|\pi/4,\; \pi/2 \right>$) one obtains the intuitive result of $\Delta\phi_0 = \pi/2$, corresponding to a quarter-waveplate, and $\alpha_0 = \pi/4$. Clearly, for any polarization state it is always possible to achieve a full dynamic range of amplitude modulation by following a suitable path in the parameter space connecting the zero of amplitude with the global maximum (cf. Fig. \ref{fig_polarization_conversion}B). However, such trajectory cannot be described with a simple analytical form and it may be convenient in practical metasurface designs to set one of the two parameters, e.g. $\Delta\phi_0$, and vary the other one to modulate the transmitted amplitude according to an analytical formula (Supplementary Material). We show in Fig. \ref{fig_polarization_conversion}F that with this restriction only a small fraction of polarization states remains limited in modulation, while for the majority it is possible to achieve a large modulation range, from 0 to almost 1.

\textit{Pure vortex generation.}---The principle of arbitrary polarization conversion illustrated above for the single nanopillar is now applied to the design of optical plates for pure vortex generation. The operation scheme is shown in Fig. \ref{fig_pure_vortex_generation}A. The input beam can be any wave of known phase and amplitude distribution, such as a plane or Gaussian wave, and arbitrary polarization. In the case of a plane wave the metasurface is designed to impart the phase and amplitude profile of an LG mode \cite{SalehTeich2007}
\begin{equation}
\begin{split}
    LG_{\ell,p} & \propto \left( \frac{r \sqrt{2}}{\omega_0} \right)^{|\ell|} L_p^{|\ell|} \left( \frac{2 r^2}{\omega_0^2} \right) \mathrm{e}^{-r^2/\omega_0^2} \mathrm{e}^{-i \ell\varphi} \\
    & \equiv A(r) \mathrm{e}^{i \psi(\varphi)}
\end{split}
\label{eq_LG_ampphase}
\end{equation}
where $\ell$ and $p$ are indices denoting the OAM and radial mode of the LG beam, $r$ and $\varphi$ are the radial and azimuthal coordinates, $\omega_0$ is the beam waist and $L_p^{|\ell|} (x)$ is the generalized Laguerre polynomial of argument $x$. The amplitude mask $A(r)$ is assimilated in the angles of the pillars $\alpha(r)$, which vary along the radial direction of the plate (inset of Fig. \ref{fig_pure_vortex_generation}A), while the azimuthal phase profile $\psi(\varphi)$ defines the pillars dimension $L_x (\varphi)$. The other pillars dimension $L_y(\varphi)$ is determined by $\psi(\varphi)+\Delta\phi_0$, where $\Delta\phi_0$ is a fixed parameter in the design corresponding to the phase retardance for the extinction condition, as previously described. The output of the metasurface is a vector vortex beam as it carries OAM and exhibits a non-uniform polarization distribution. After the beam is filtered by a linear polarizer, which could be a separate element or directly integrated in the metasurface substrate as a wire-grid polarizer \cite{Divitt2019}, its amplitude distribution corresponds to that of a pure LG beam.

The near field intensity maps of an $LG_{5,0}$ vortex generator obtained by FDTD simulations are shown in Fig. \ref{fig_pure_vortex_generation}B. The uniform intensity of the input plane wave is separated by the metasurface into a vertically-polarized component, which is aligned with the polarizer axis and exhibits the characteristic ring-shaped distribution of a vortex mode, and a horizontally-polarized component, which contains essentially the complementary intensity distribution and will be filtered by the polarizer. (A small fraction of light goes also into the longitudinal component, which is not shown here.) After propagating to the far field the component filtered by the polarizer we obtain the intensity and phase distributions shown in Fig. \ref{fig_pure_vortex_generation}D, which can be compared with those obtained from a PO metasurface operating without a polarizer (Fig. \ref{fig_pure_vortex_generation}C). While in both cases the phase shows an azimuthal modulation consisting of five sectors ($\ell=5$), only the PA metasurface produces a single ring of intensity as one would expect for the fundamental radial mode ($p=0$). The PO metasurface instead shows multiple rings corresponding to the superposition of different $p$-modes \cite{Sephton:16}. A modal purity analysis shows that 97$\%$ of the generated beam power is in the $p=0$ mode for the PA metasurface, while this value drops to 23$\%$ for the PO metasurface. The efficiency in generating pure beams with the two types of metasurfaces will be discussed in the next section.

The purity of the intensity profile generated by the PA metasurface all originates from a proper use of the pillar angle degree of freedom, as highlighted by the comparison of the device designs (cf. Fig \ref{fig_pure_vortex_generation}C and \ref{fig_pure_vortex_generation}D). As described in our theory, the PA control works for any input polarization state. In Fig. \ref{fig_pure_vortex_generation}E we show that also for elliptically polarized light, where we chose as an example $\left|\chi=\pi/5, \; \delta = \pi/3 \right>$, one can generate a high purity vortex mode with a single intensity ring. While the generation of such beams is of interest in OAM applications where the energy needs to be confined in the fundamental radial mode, the control of higher-order $p$-modes may be appealing in other applications, such as super-resolution microscopy. In Fig. \ref{fig_pure_vortex_generation}F we show that this is possible using a PA metasurface, demonstrating as an example the generation of an $LG_{1,3}$ mode. Also in this case nearly all (92$\%$) the generated beam power is retained in the target mode.

\begin{figure}[t]
    \centering
    \includegraphics[width=0.45\textwidth]{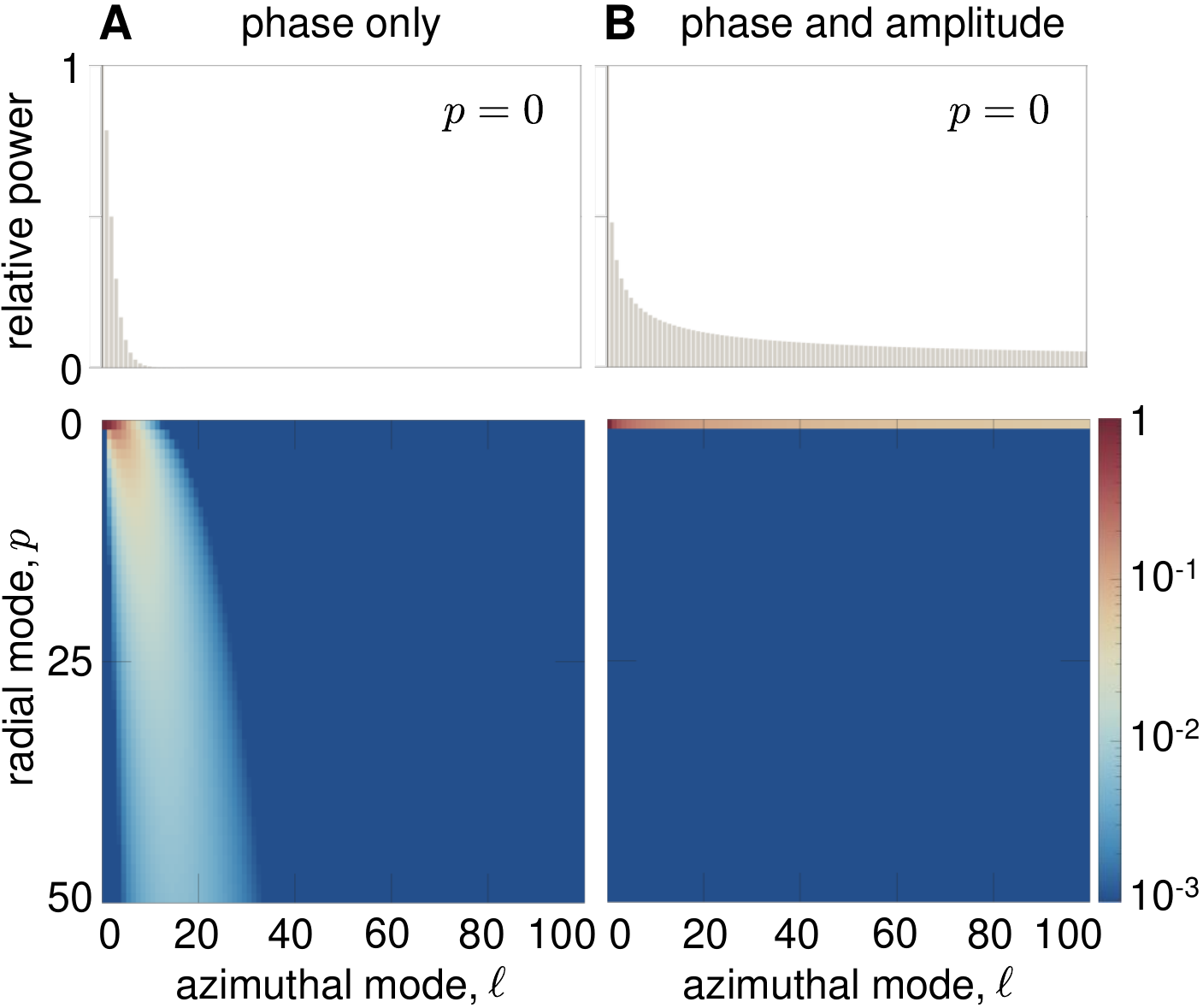}
    \caption{\textbf{Purity efficiency of phase-only and phase-amplitude transformations.} Fraction of the input beam power that is contained in the fundamental radial mode $p=0$ (top) and distributed among all $p$-modes (bottom) for different values of orbital angular momentum $\ell$ imparted by a metasurface plate. Both the case of phase-only metasurfaces (\textbf{A}) and phase-amplitude metasurfaces (\textbf{B}) are shown.}
    \label{fig_efficiency}
\end{figure}

\textit{Discussion.}---After having demonstrated that PA metasurfaces perform better than PO metasurfaces in terms of pure vortex generation, we wish to compare their efficiencies. We define the efficiency as the fraction of power of the beam incident on the metasurface that is converted into the target $LG_{\ell,0}$ mode. For this comparison it is important to consider an input wave that carries finite power and that can be readily available as a practical source in a laboratory, thus we choose a Gaussian beam. In the case of PO metasurfaces the efficiency is known to drop rapidly with the OAM charge \cite{Sephton:16}. As shown in Fig. \ref{fig_efficiency}A, the converted beam power spreads towards higher radial modes as $|\ell|$ grows, due to the lack of the amplitude modulation term, and nearly no energy is contained in the $p=0$ mode for $|\ell|>10$. In PA metasurfaces instead, all the power transmitted through the polarizer is already in the $p=0$ mode, thus no modal filtering is needed. What limits the efficiency in this case is the fraction of input power that needs to be absorbed by the polarizer. To maximize the generation efficiency there is an optimum choice for the beam waist $\omega_S$ of the Gaussian source, which is in general larger than the beam waist $\omega_0$ of the Gaussian embedded in the target $LG$ mode (Eq. \ref{eq_LG_ampphase}) and is calculated as $\omega_S/\omega_0=\sqrt{|\ell|+1}$ based on the overlap of the source intensity profile with the transmittance mask of the metasurface (Supplementary Material). The efficiency of the PA metasurface for the optimum $\omega_S/\omega_0$ ratio is shown in Fig. \ref{fig_efficiency}B. It shows a clear benefit for the generation of beams with large OAM charge, as it remains sizeable even for very large values of $|\ell|$ (e.g. 5$\%$ efficiency at $|\ell|=100$). Another advantage of PA metasurfaces in this respect lies in the sampling of the azimuthal phase profile. While PO metasurfaces suffer from poor phase resolution close to the singularity point, due to the strong phase gradient of large OAM beams and finite size of the meta-atoms, in the case of PA metasurfaces the phase sampling only matters in the ring-shaped intensity region of the target $LG_{\ell,0}$ mode. Considering that the peak intensity of the ring occurs at a radial position $r_\mathrm{max} = \omega_0 \sqrt{|\ell|/2}$ from the singularity and that the length of an arc spanning a phase period from 0 to $2\pi$ is $p=2\pi\:r_\mathrm{max}/|\ell|$, we can estimate the number of pillars covering a phase period in a PA metasurface as
\begin{equation}
    \frac{p}{U} \sim \sqrt{\frac{2}{|\ell|}} \frac{\pi \omega_0}{U}
\end{equation}
where $U$ is the size of the unit cell of the metasurface. Thus for any OAM charge and for a unit cell size fixed by the phase library, the phase sampling of the PA metasurface can be chosen arbitrarily large just by adjusting $\omega_0$.

The PA metasurfaces introduced here also present some limitations with respect to the existing approaches for optical vortex generation. Differently from $J$-plates \cite{Devlin2017,Huang2019}, which implement PO trasformations, our PA metasurfaces are not spin-orbit converters. In particular, they can operate only on a single input polarization state. The degree of freedom that is used in $J$-plates to achieve independent OAM conversion of two orthogonal input polarization states, here is used to apply a desired amplitude modulation. In comparison with $q$-plate lasers \cite{Maguid2018} and $J$-plate lasers \cite{Sroor2020}, which incorporate PO transformation optics in active resonator cavities, our PA metasurfaces cannot convert all the input energy into a pure LG state nor they can generate a continuously-variable superposition of OAM states \cite{Sroor2020}, only a fixed one. In view of these differences we do not expect that PA metasurfaces will replace the existing technologies for OAM beam generation but may represent a very convenient and powerful approach in several applications, thanks to their simple and compact implementation scheme, sizeable efficiency and high purity.

\bigskip
This work has been financially supported by the European Research Council (ERC) under the European Union’s Horizon 2020 research and innovation programme ``METAmorphoses'', grant agreement no. 817794.

\bigskip

\bibliography{pure_vortex_bib}

\end{document}


 
  \title{Supplementary Material to:\\
  Arbitrary polarization conversion for pure vortex generation with a single metasurface}
 \author{Marco Piccardo}
  \email{marco.piccardo@iit.it}
 \affiliation{Center for Nano Science and Technology,\\ Fondazione Istituto Italiano di Tecnologia, Milan 20133, Italy}

 \author{Antonio Ambrosio}
 \email{antonio.ambrosio@iit.it}
 \affiliation{Center for Nano Science and Technology,\\ Fondazione Istituto Italiano di Tecnologia, Milan 20133, Italy}

\maketitle

\textit{Extinction conditions.}---In the main text we solved the general problem corresponding to the determination of the pillar parameters that provide complete attenuation of an input field of arbitrary polarization passing through the pillar-polarizer system. The expression for the pillar orientation angle $\alpha_0$ for the extinction condition was given in Eq. (3) of the main text. The analytical expression for the phase retardance $\Delta\phi_0$ is less compact and is given here:
\begin{equation}
\begin{split}
\Delta\phi_0 = & 2 \; \Big\{  -\mathrm{cot}^{-1} \Big[ \mathrm{csc} \: \delta \Big( \mathrm{cos} \: \delta + \mathrm{cot} \: \alpha_0 \; \mathrm{cot} \: \chi + \mathrm{csc} \: \alpha_0 \; \mathrm{csc} \: \chi \\
& \sqrt{\mathrm{sin}^2 \: \alpha_0 \Big( \Big( \mathrm{cos} \: \delta + \mathrm{cot} \: \alpha_0 \; \mathrm{cot} \: \chi \Big)^2 + \mathrm{sin}^2 \delta \Big) \: \mathrm{sin}^2 \: \chi  } \; \bigg) \bigg] \\
& + \mathrm{cot}^{-1} \Big[ \mathrm{csc} \: \delta \Big( \mathrm{cos} \: \delta - \mathrm{cot} \: \chi \; \mathrm{tan} \: \alpha_0 + \mathrm{csc} \: \chi \; \mathrm{sec} \: \alpha_0 \\
& \sqrt{ \mathrm{cos}^2 \: \alpha_0 \; \mathrm{sin}^2 \: \chi \Big( \mathrm{sin}^2 \delta + \Big(  \mathrm{cos} \: \delta - \mathrm{cot} \: \chi \; \mathrm{tan} \: \alpha_0 \Big)^2 \Big) } \; \bigg) \bigg] \bigg\}
\end{split}
\label{eq_deltaphi0}
\end{equation}
where $\left|\chi,\; \delta \right>$ characterize the incident polarization state (Eq. (2) of the main text).

\begin{figure}[t!]
    \centering
    \includegraphics[width=0.9\textwidth]{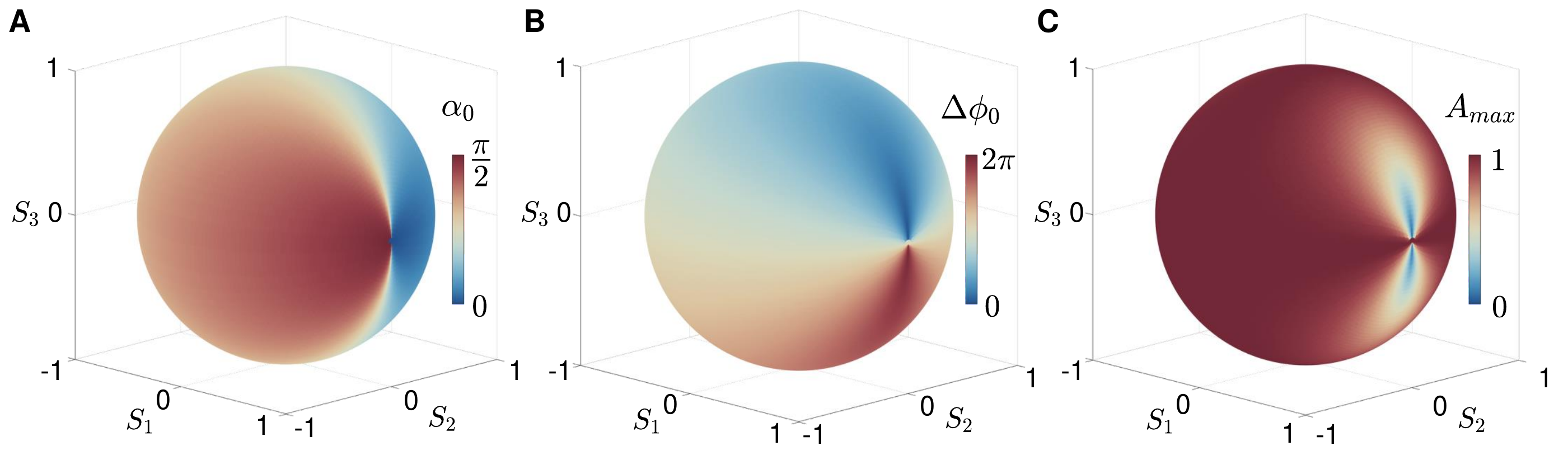}
    \caption{Representation on the Poincaré sphere of the design parameters required for the extinction condition through the pillar-polarizer system, corresponding to the pillar angle (\textbf{A}) and phase retardance (\textbf{B}), for any input polarization state. \textbf{C}, Maximum field transmission through the system when the phase retardance is fixed to $\Delta\phi_0$ and the orientation angle is allowed to vary. These plots correspond to the planar representations shown in Fig. 2 D-F of the main text. The axis are the three Stokes parameters, $S_1$, $S_2$ and $S_3$.}
    \label{figS_Poincare}
\end{figure}

The dependence of $\alpha_0$ and $\Delta\phi_0$ were plotted in the main text on the $\left( \chi, \delta \right)$ plane. In Fig. \ref{figS_Poincare} we represent them on the Poincaré sphere, together with the distribution of maximum field transmission for the case in which the phase retardance is fixed to $\Delta\phi_0$, as in Eq. (\ref{eq_deltaphi0}), and the orientation angle is allowed to vary to maximize the transmission. Only a limited fraction of input states results in a small transmission. Such states are oriented close to the $x$-axis, almost perpendicularly to the $y$-oriented polarizer and therefore nearly no amplitude can be transmitted, regardless of the angle of the pillar. The maximum transmitted amplitude on the Poincaré sphere shows a distribution that is reminiscent of that of a dipole emission pattern (Fig. \ref{figS_Poincare}C). This can be understood as in our system, similarly to a dipole, there is a preferential direction (in our case determined by the polarizer axis) from which no amplitude can be emitted. However we note that, as already mentioned in the main text, even for such polarization states it is possible to obtain unity transmission through the pillar-polarizer system if both the orientation angle and the phase retardance of the pillar are allowed to vary. In fact it can be shown that for any input polarization state a combination of these parameters always exist such to provide unity transmission (cf. Fig. 1B of the main text). The only disadvantage of the approach where both the phase retardance and the angle can vary is that the parameter combination for an arbitrary amplitude modulation needs to be determined numerically, rather than analytically.

\begin{figure}[t]
    \centering
    \includegraphics[width=0.3\textwidth]{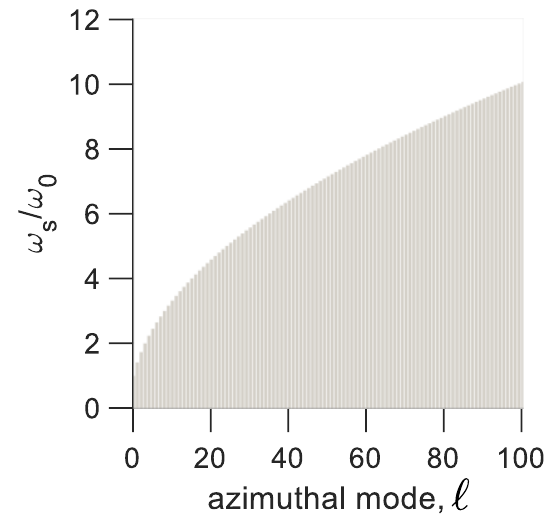}
    \caption{Ratio of the beam waist of a Gaussian source and beam waist of the embedded Gaussian of a Laguerre-Gauss mode generated by the PA metasurface, such that the purity conversion efficiency is maximized. The ratio is plotted for different orbital angular momentum charges of the generated Laguerre-Gauss mode.}
    \label{figS_wsw0}
\end{figure}

\textit{Purity efficiency.}---The efficiency of the PA metasurface in converting an input beam into a pure Laguerre-Gauss (LG) mode depends on the radial distribution of both the source and LG fields, which are given by
\begin{equation}
    E_S = e^{- (r/\omega_S)^2}
    \label{eq_ES}
\end{equation}
and
\begin{equation}
    E_{LG} = e^{- (r/\omega_0)^2} \; \left( \frac{r \sqrt{2}}{\omega_0} \right)^{|\ell|},
    \label{eq_ELG}
\end{equation}
respectively, where $r$ is the radial coordinate, $\omega_S$ is the beam waist of the Gaussian source and $\omega_0$ is the beam waist of the embedded Gaussian in the LG mde. In Eq. (\ref{eq_ES}) and (\ref{eq_ELG}) we omitted any constant or azimuthally-varying factor. The purity efficiency of the PA metasurface in terms of power conversion is then defined as
\begin{equation}
    \eta = \frac{\int\limits_0^\infty E^2_{LG} \mathrm{d}r}{\mathcal{N}^2 \; \int\limits_0^\infty E^2_{S} \mathrm{d}r}
    \label{eq_eta}
\end{equation}
where $\mathcal{N}$ is a normalization constant corresponding to the maximum of $E_{LG}/E_S$. By maximizing Eq. (\ref{eq_eta}) vs. $\omega_S$ it is calculated that the maximum efficiency for a given $|\ell|$ is obtained when $\omega_S/\omega_0 = \sqrt{|\ell| + 1}$ (Fig. \ref{figS_wsw0}). In these conditions the purity efficiency can be written as
\begin{equation}
    \eta = e^{|\ell|} \left[ \frac{\omega_S}{\omega_0} \sqrt{\frac{|\ell|}{(\omega_S/\omega_0)^2 - 1}} \right]^{-2|\ell|} \; \Gamma\left(  \frac{1}{2} + |\ell| \right) \; \frac{\omega_0}{\sqrt{\pi} \: \omega_S}
\end{equation}
where $\Gamma(...)$ is the gamma function.

\begin{figure}
\centering
    \includegraphics[width=0.9\textwidth]{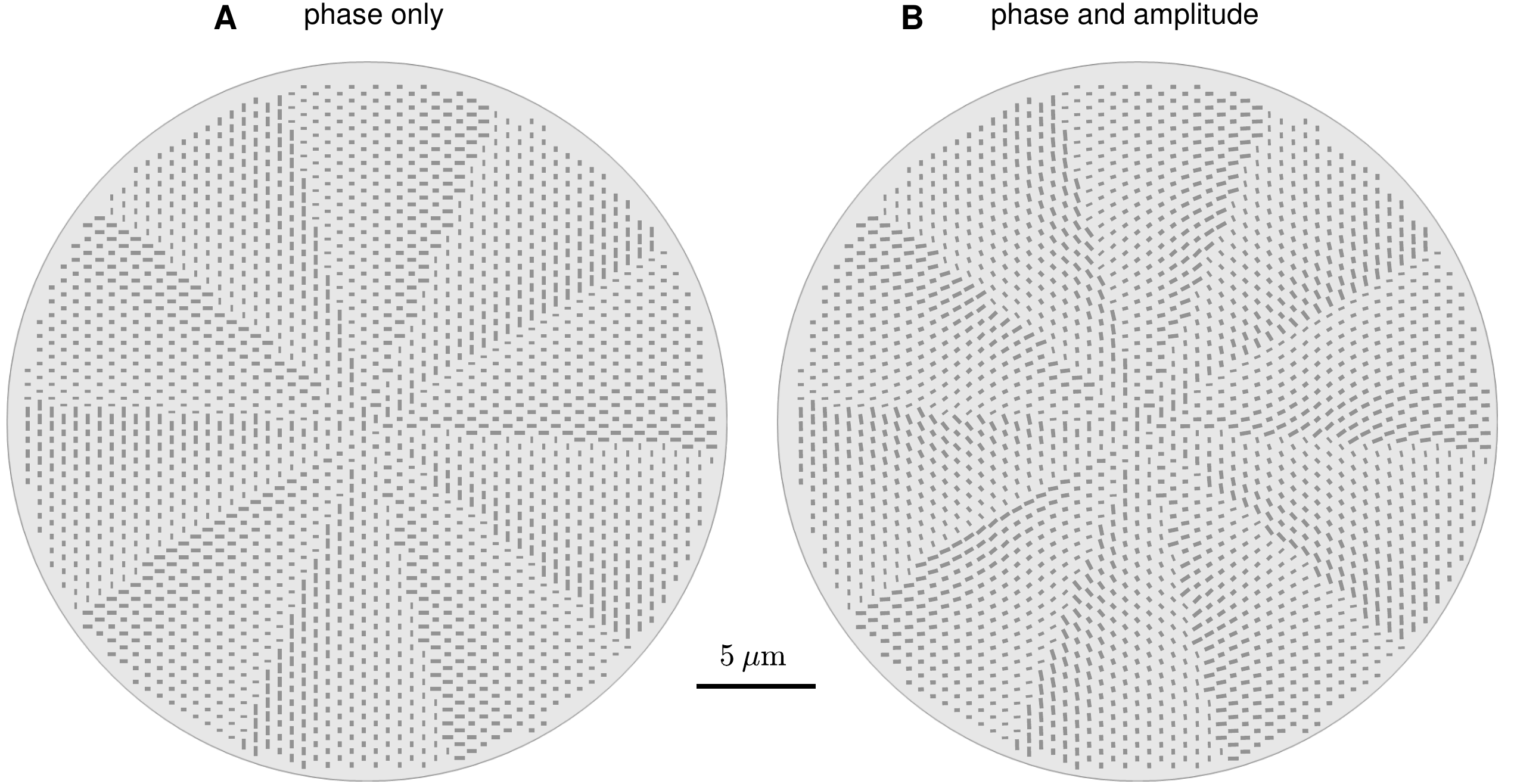}
    \caption{Comparison of the device designs for a PO and PA metasurface imparting an orbital angular momentum charge of $|\ell|=5$. Only the PA metasurface incorporates an amplitude transmission mask, which is embedded in the pillars rotation.}
\label{figS_POPAdesign}
\end{figure}

\textit{Device design.}---In Fig. 2 of the main text only the central regions of the designed metasurfaces are shown due to space constraints. In Fig. \ref{figS_POPAdesign}A,B we show a comparison between the PO and PA metasurface designs on a larger scale. In both cases the target LG mode corresponds to an azimuthal mode with $|\ell|=5$. It can be appreciated as the angle degree of freedom is used in the PA metasurface for the purpose of implementing the amplitude mask, which is essential for the generation of pure LG modes.